\documentclass[twocolumn,aps,prl]{revtex4}

\usepackage{graphicx}
\usepackage{amsmath}

\begin{document}

\title{
Approaching Unit Visibility for Control of a Superconducting
Qubit\\ with Dispersive Readout}

\author{A.~Wallraff}
\affiliation{Departments of Applied Physics and Physics, Yale
University, New Haven, CT 06520}
\author{D.~I.~Schuster}
\affiliation{Departments of Applied Physics and Physics, Yale
University, New Haven, CT 06520}
\author{A.~Blais}
\affiliation{Departments of Applied Physics and Physics, Yale
University, New Haven, CT 06520}
\author{L.~Frunzio}
\affiliation{Departments of Applied Physics and Physics, Yale
University, New Haven, CT 06520}
\author{J.~Majer}
\affiliation{Departments of Applied Physics and Physics, Yale
University, New Haven, CT 06520}
\author{S.~M.~Girvin}
\affiliation{Departments of Applied Physics and Physics, Yale
University, New Haven, CT 06520}
\author{R.~J.~Schoelkopf}
\affiliation{Departments of Applied Physics and Physics, Yale
University, New Haven, CT 06520}
\date{\today}

\begin{abstract}
In a Rabi oscillation experiment with a superconducting qubit we
show that a visibility in the qubit excited state population of
more than $90 \, \%$ can be attained. We perform a dispersive
measurement of the qubit state by coupling the qubit
non-resonantly to a transmission line resonator and probing the
resonator transmission spectrum. The measurement process is well
characterized and quantitatively understood. The qubit coherence
time is determined to be larger than $500 \, \rm{ns}$ in a
measurement of Ramsey fringes.
\end{abstract}

\maketitle

One of the most promising solid-state architectures for the
realization of a quantum information processor \cite{Nielsen00} is
based on superconducting electrical circuits \cite{Devoret04}. A
variety of such circuits acting as qubits \cite{Nielsen00}, the
basic carriers of quantum information in a quantum computer, have
been created and their coherent control has been demonstrated
\cite{Nakamura99,Vion02,Martinis02,Yu02,Chiorescu03,Collin04}.
Recent experiments have realized controlled coupling between
different qubits
\cite{Berkley03,Pashkin03,Majer04,Chiorescu04,McDermott04} and
also first two-qubit quantum logic gates \cite{Yamamoto03}.

An outstanding question for superconducting qubits, and in fact
for all solid-state implementations of quantum information
processors, is whether the qubits are sufficiently well-isolated
to allow long coherence times and high-fidelity preparation and
control of their quantum states. This question is complicated by
the inevitable imperfections in the measurement. A canonical
example is a Rabi oscillation experiment, where the experimenter
records the oscillations of a meter's response as a function of
pulse length to infer the qubit's excited state population
immediately after the pulse. The measurement contrast (e.g.~the
amplitude of the meter's measured swing relative to its maximum
value) is reduced in general by both errors in the qubit
preparation and readout, and sets only a lower limit on the
visibility of oscillations in the qubit population. Most
experiments with superconducting qubits to date have reported only
the measurement contrast, typically in the range of $10-50\,\%$
\cite{Nakamura99,Vion02,Martinis02,Yu02,Chiorescu03,Collin04,Yamamoto03}.

A full understanding of the measurement process is required to
extract the qubit population from the meter's output. The qubit
control and read-out is then characterized by the visibility,
defined as the maximum qubit population difference observed in a
Rabi oscillation or Ramsey fringe experiment. Moreover, it is
essential to demonstrate that a qubit can be controlled without
inducing undesired leakage to other qubit states or entanglement
with the environment. We note that experiments have suggested that
there can be substantial reduction of the visibility due to
entanglement with spurious environmental fluctuators
\cite{Simmonds03}.

In this letter, we report results on time-domain control of the
quantum state of a superconducting qubit, where the qubit state is
measured using a dispersive microwave measurement in a circuit
quantum electrodynamics (QED) architecture \cite{Blais04}. This
novel technique has shown good agreement with predictions in
steady-state experiments \cite{schuster05}. Here, we observe the
measurement response, both during and after qubit state
manipulation, which is in quantitative agreement with the
theoretical model of the system, allowing us to separate the
contributions of the qubit and the readout to the observed
contrast. The observed contrast of $85\,\%$ and a visibility of
greater than $95\,\%$ for Rabi oscillations demonstrates that high
accuracy control can be achieved in superconducting qubits.

\begin{figure}[!bp]
\includegraphics[width = 0.85 \columnwidth]{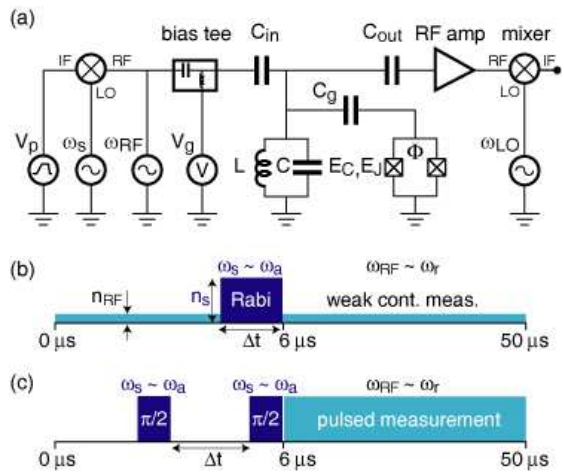}
\caption{(color online) (a) Simplified circuit diagram of
measurement setup. A Cooper pair box with charging energy
$E_{\rm{C}}$ and Josephson energy $E_{\rm{J}}$ is coupled through
capacitor $C_{\rm{g}}$ to a transmission line resonator, modelled
as parallel combination of an inductor $L$ and a capacitor $C$.
Its state is determined in a phase sensitive heterodyne
measurement of a microwave transmitted at frequency
$\omega_{\rm{RF}}$ through the circuit, amplified and mixed with a
local oscillator at frequency $\omega_{\rm{LO}}$. The Cooper pair
box level separation is controlled by the gate voltage $V_g$ and
flux $\Phi$. Its state is coherently manipulated using microwaves
at frequency $\omega_{\rm{s}}$ with pulse shapes determined by
$V_{\rm{p}}$ \cite{Collin04}. (b) Measurement sequence for Rabi
oscillations with Rabi pulse length $\Delta t$, pulse frequency
$\omega_{\rm{s}}$ and amplitude $\propto \sqrt{n_{\rm{s}}}$ with
continuous measurement at frequency $\omega_{\rm{RF}}$ and
amplitude $\propto \sqrt{n_{\rm{RF}}}$. (c) Sequence for Ramsey
fringe experiment with two $\pi/2$-pulses at $\omega_{\rm{s}}$
separated by a delay $\Delta t$ and followed by a pulsed
measurement.} \label{fig:setup}
\end{figure}

In our circuit QED architecture \cite{Blais04}, a Cooper pair box
\cite{Bouchiat98}, acting as a two level system with ground
$\left|\downarrow\right\rangle$ and excited states
$\left|\uparrow\right\rangle$ and level separation $E_a = \hbar
\omega_a = \sqrt{E_{\rm{el}}^2+E_{\rm{J}}^2}$ is coupled
capacitively to a single mode of the electromagnetic field of a
transmission line resonator with resonance frequency
$\omega_{\rm{r}}$, see Fig.~\ref{fig:setup}a. As demonstrated for
this system, the electrostatic energy $E_{\rm{el}}$ and the
Josephson energy $E_{\rm{J}}$ of the split Cooper pair box can be
controlled \textsl{in situ} by a gate voltage $V_g$ and magnetic
flux $\Phi$ \cite{wallraff04c,schuster05}, see
Fig.~\ref{fig:setup}a. In the resonant ($\omega_a=\omega_r$)
strong coupling regime a single excitation is exchanged coherently
between the Cooper pair box and the resonator at a rate $g/\pi$,
also called the vacuum Rabi frequency \cite{wallraff04c}. In the
non-resonant regime ($\left|\Delta\right| =
\left|\omega_a-\omega_r\right| > g$) the capacitive interaction
gives rise to a dispersive shift $(g^2/\Delta)\sigma_z$ in the
resonance frequency of the cavity which depends on the qubit state
$\sigma_z$, the coupling $g$ and the detuning $\Delta$
\cite{Blais04,schuster05}. We have suggested that this shift in
resonance frequency can be used to perform a quantum
non-demolition (QND) measurement of the qubit state
\cite{Blais04}. With this technique we have recently measured the
ground state response and the excitation spectrum of a Cooper pair
box \cite{wallraff04c,schuster05}.

\begin{figure}[!bp]
\includegraphics[width = 0.85 \columnwidth]{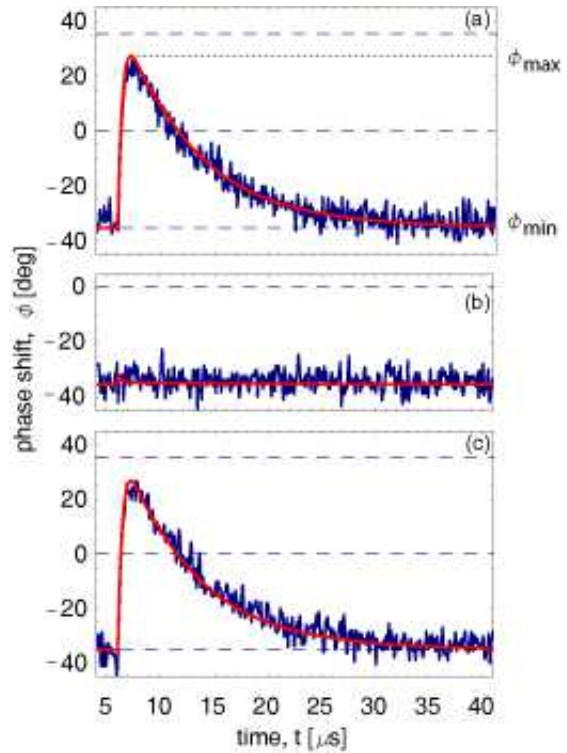}
\caption{(color online) Measurement response $\phi$ (blue lines)
and theoretical prediction (red lines) \textsl{vs.}~time. At
$t=6\,\rm{\mu s}$ (a) a $\pi$ pulse, (b) a $2\pi$ pulse, and (c) a
$3\pi$ pulse is applied to the qubit. In each panel the dashed
lines correspond to the expected measurement response in the
ground state $\phi_{\left|\downarrow\right\rangle}$, in the
saturated state $\phi=0$, and in the excited state
$\phi_{\left|\uparrow\right\rangle}$.} \label{fig:piresponse}
\end{figure}

In the experiments presented here, we coherently control the
quantum state of a Cooper pair box by applying to the qubit
microwave pulses of frequency $\omega_s$, which are resonant with
the qubit transition frequency $\omega_{\rm{a}}/2\pi \approx 4.3
\, \rm{GHz}$, through the input port $C_{\rm{in}}$ of the
resonator, see Fig.~\ref{fig:setup}a. The microwaves drive Rabi
oscillations in the qubit at a frequency of $\nu_{\rm{Rabi}} =
\sqrt{n_{\rm{s}}}g/\pi$, where $n_{\rm{s}}$ is the average number
of drive photons \emph{within} the resonator. Simultaneously, we
perform a continuous dispersive measurement of the qubit state by
determining both the phase and the amplitude of a coherent
microwave beam of frequency $\omega_{\rm{RF}}/2\pi =
\omega_{\rm{r}}/2\pi \approx 5.4 \, \rm{GHz}$ transmitted through
the resonator \cite{Blais04,wallraff04c}. The phase shift $\phi =
\tan^{-1}\left(2g^2/\kappa\Delta\right)\sigma_z$ is the response
of our meter from which we determine the qubit population. For the
measurement, we chose a resonator that has a quality factor of $Q
\sim 0.7\times10^4$ corresponding to a photon decay rate of
$\kappa/2\pi = 0.73\,\rm{MHz}$. The resonator is populated with $n
\sim 1$ measurement photons on average, where $n$ is calibrated
using the ac-Stark shift \cite{schuster05}.

We initially determine the maximum swing of the meter in a
calibration measurement by first maximizing the detuning $\Delta$
to minimize the interaction ($g^2/\Delta \rightarrow 0$) which
defines $\phi = 0$. Then, we prepare the Cooper pair box in the
ground state $\left|\downarrow\right\rangle$ and bias it at charge
degeneracy where the detuning is adjusted to $ \Delta/2\pi \approx
- 1.1 \, \rm{GHz}$, corresponding to a maximum in the Josephson
coupling energy of $E_{\rm{J}}/h\approx4.3\,\rm{GHz} <
\omega_{\rm{r}}/2\pi$ for this particular sample. In this case we
measure a minimum meter response of
$\phi_{\left|\downarrow\right\rangle} = - 35.3 \, \rm{deg}$
corresponding to a coupling strength of $g/2\pi =17 \, \rm{MHz}$.
Saturating the qubit transition by applying a long microwave pulse
which incoherently mixes the ground and excited states such that
the occupation probabilities are
$P_{\left|\downarrow\right\rangle} =
P_{\left|\uparrow\right\rangle} = 1/2$, the measured phase shift
is found to be $\phi=0$, as expected \cite{schuster05}. From these
measurements, the predicted phase shift in the measurement beam
induced by a fully polarized qubit
($P_{\left|\uparrow\right\rangle} = 1$) would be
$\phi_{\left|\uparrow\right\rangle} = 35.3 \, \rm{deg}$. Thus, the
maximum swing of the meter is bounded by
$\phi_{\left|\uparrow\right\rangle}-\phi_{\left|\downarrow\right\rangle}$.

In our continuous measurement of Rabi oscillations, the qubit is
initially prepared in the ground state by relaxation, the thermal
population of excited states being negligible. Then, a short
microwave pulse of length $\Delta t$ prepares the qubit state
while the measurement response $\phi$ is continuously monitored.
This sequence is repeated every $50 \, \rm{\mu s}$, see
Fig.~\ref{fig:setup}b, and the measurement response is digitally
averaged $5\times10^4$ times. The signal to noise ratio (SNR) in
the averaged value of $\phi$ in an integration time of $100 \,
\rm{ns}$ is approximately $25$, see Fig.~\ref{fig:piresponse},
corresponding to a SNR of $0.1$ in a single shot. For the present
experimental setup the single shot read-out fidelity for the qubit
state integrated over the relaxation time ($T_1\sim 7 \rm{\mu s}$)
is approximately $30 \, \%$ \cite{Schuster05b}. Both, a read-out
amplifier with lower noise temperature or a larger signal power
would potentially allow to realize high fidelity single shot
measurements of the qubit state in this setup.

The time dependence of the averaged value of $\phi$ in response to
a $\pi$ pulse of duration $\Delta t = 1/2\nu_{\rm{Rabi}} \sim 16
\, \rm{ns}$ applied to the qubit is shown in
Fig.~\ref{fig:piresponse}a. Before the start of the pulse the
measured phase shift is observed to be
$\phi_{\left|\downarrow\right\rangle} \approx - 35.3 \, \rm{deg}$
corresponding to the qubit being in the ground state. Due to the
state change of the qubit induced by the pulse, the resonator
frequency is pulled by $2g^2/\Delta$ and, thus, the measured phase
shift is seen to rise exponentially towards
$\phi_{\left|\uparrow\right\rangle}$ on the time scale of the
intrinsic response time of the measurement  $2/\kappa \approx 400
\, \rm{ns}$, i.e.~twice the photon life time. At the end of the
$\pi$ pulse, the qubit excited state decays exponentially on the
time scale of its energy relaxation time $T_1 \sim 7.3 \, \rm{\mu
s}$, as extracted from the decay in the measured phase shift, see
Fig.~\ref{fig:piresponse}a. As a result, the maximum measured
response $\phi_{\rm{max}}$ does not reach the full value of
$\phi_{\left|\uparrow\right\rangle}$. In general, the measurement
contrast $C = (\phi_{\rm{max}}- \phi_{\rm{min}})
/(\phi_{\left|\uparrow\right\rangle}-\phi_{\left|\downarrow\right\rangle})$
will be reduced in any qubit read-out for which the timescale of
the qubit decay is not infinitely longer than the response time of
the measurement. Additionally, in non-QND measurements the
contrast is reduced even further due to mixing of the qubit states
induced by the interaction with the measurement appartus. In our
QND measurement presented here, the qubit decay time is about 15
times longer than the response time of the measurement, allowing
us to reach a high maximum contrast of $C \sim 85 \,\%$ in the
bare measurement response $\phi$.

\begin{figure}[bp]
\includegraphics[width = 0.70 \columnwidth]{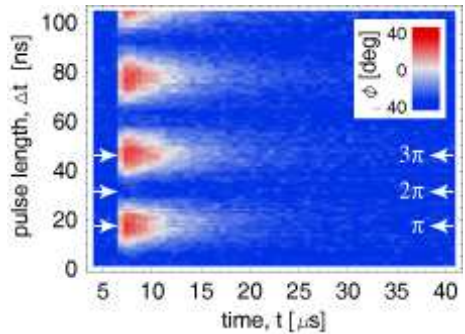}
\caption{(color online) Color density plot of phase shift $\phi$
(see inset for scale) versus measurement time $t$ and Rabi pulse
length $\Delta t$. Data shown in Fig.~\ref{fig:piresponse} are
slices through this data set at the indicated pulse lengths.}
\label{fig:2DRabi}
\end{figure}

In Figs.~\ref{fig:piresponse}b and c, the measured response $\phi$
of the meter to a $2\pi$ and a $3\pi$ pulse acting on the qubit is
shown. As expected, no phase shift is observable for the $2\pi$
pulse since the response time of the resonator read-out is much
slower than the duration $\Delta t = 32 \, \rm{ns}$ of the pulse.
In agreement with the expectations for this QND scheme, the
measurement does not excite the qubit out of its ground state,
i.e.~$\phi_{\rm{min}} = \phi_{\rm{max}} =
\phi_{\left|\downarrow\right\rangle}$. The response to the $3 \pi$
pulse is virtually indistinguishable from the one to the $\pi$
pulse, as expected for the long coherence and energy relaxation
times of the qubit. In the 2D density plot Fig.~\ref{fig:2DRabi},
Rabi oscillations are clearly observed in the phase shift acquired
versus measurement time $t$ and Rabi pulse length $\Delta t$.

The observed measurement response $\phi$ is in excellent agreement
with the theoretical predictions, see red lines in
Fig.~\ref{fig:piresponse}, demonstrating a good understanding of
the measurement process. The temporal response $\phi(t) = \arg \{i
\langle a(t)\rangle\}$ of the cavity field $a$ is calculated by
deriving and solving Bloch-type equations of motion for the cavity
and qubit operators \cite{Blais05b} using the Jaynes-Cummings
Hamiltonian in the dispersive regime \cite{Blais04,schuster05} as
the starting point. A semi-classical factorization approximation
is done to truncate the resulting infinite set of equations to a
finite set (e.g. $\langle a^\dag a \sigma_z \rangle\sim\langle
a^\dag a \rangle\langle \sigma_z \rangle$; all lower order
products are kept). This amounts to neglecting higher order
correlations between qubit and field which is a valid
approximation in the present experiment. The calculations
accurately model the exponential rise in the observed phase shift
on the time scale of the resonator response time due to a state
change of the qubit. They also accurately capture the reduced
maximum response $\phi_{\rm{max}}$ due to the exponential decay of
the qubit. Overall, excellent agreement in the temporal response
of the measurement is found over the full range of relevant qubit
and measurement time scales, see Fig.~\ref{fig:piresponse}. All
parameters used in the comparison of theory with experimental data
have been extracted from independent measurements. The parameters
remain fixed for all Rabi pulse length and thus leave no
adjustables.

\begin{figure}[!bp]
\includegraphics[width = 0.95 \columnwidth]{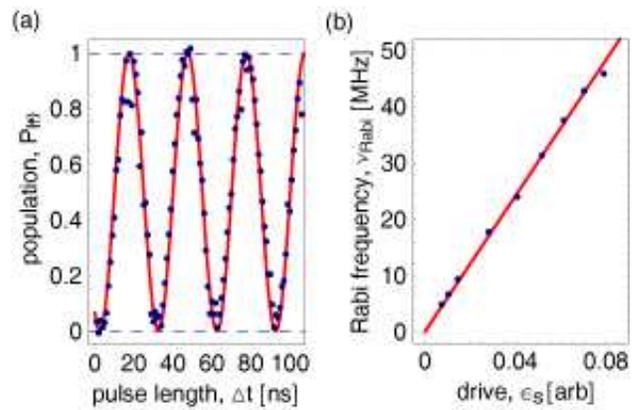}
\caption{(color online) (a) Rabi oscillations in the qubit
population $P_{\left|\uparrow\right\rangle}$ \textsl{vs.} Rabi
pulse length $\Delta t$ (blue dots) and fit with unit visibility
(red line). (b) Measured Rabi frequency $\nu_{\rm{Rabi}}$
\textsl{vs.} pulse amplitude $\epsilon_{\rm{s}}$ (blue dots) and
linear fit.} \label{fig:rabioscillations}
\end{figure}

The visibility of the excited state population
$P_{\left|\uparrow\right\rangle}$ in the Rabi oscillations is
extracted from the time dependent measurement response $\phi$ for
each Rabi pulse length $\Delta t$. We find
$P_{\left|\uparrow\right\rangle}$ by calculating the normalized
dot product between the response $\phi$ and the theoretically
predicted response taking into account the systematics of the
measurement. This amounts to comparing the full area under a
measured response curve, such as those shown in
Fig.~\ref{fig:piresponse}, to the theoretically predicted area.
The averaged response of all measurements taken over a window in
time extending from the start of the Rabi pulse out to several
qubit decay times $T_1$ is used in our continuous dispersive
measurement to extract $P_{\left|\uparrow\right\rangle}$. This
maximizes the signal to noise ratio in the extracted Rabi
oscillations.

The extracted qubit population $P_{\left|\uparrow\right\rangle}$
is plotted versus $\Delta t$ in Fig.~\ref{fig:rabioscillations}a.
We observe a visibility of $95 \pm 6 \,\%$ in the Rabi
oscillations with error margins determined from the residuals of
the experimental $P_{\left|\uparrow\right\rangle}$ with respect to
the predicted values. Thus, in a measurement of Rabi oscillations
in a superconducting qubit, a visibility in the population of the
qubit excited state that approaches unity is observed for the
first time. Moreover, we note that the decay in the Rabi
oscillation amplitude out to pulse lengths of $100 \, \rm{ns}$ is
very small and consistent with the long $T_1$ and $T_2$ times of
this charge qubit, see Fig.~\ref{fig:rabioscillations}a and Ramsey
experiment discussed below. We have also verified the expected
linear scaling of the Rabi oscillation frequency $\nu_{\rm{Rabi}}$
with the pulse amplitude
$\epsilon_{\rm{s}}\propto\sqrt{n_{\rm{s}}}$, see
Fig.~\ref{fig:rabioscillations}b.

\begin{figure}[!t]
\includegraphics[width = 1.0 \columnwidth]{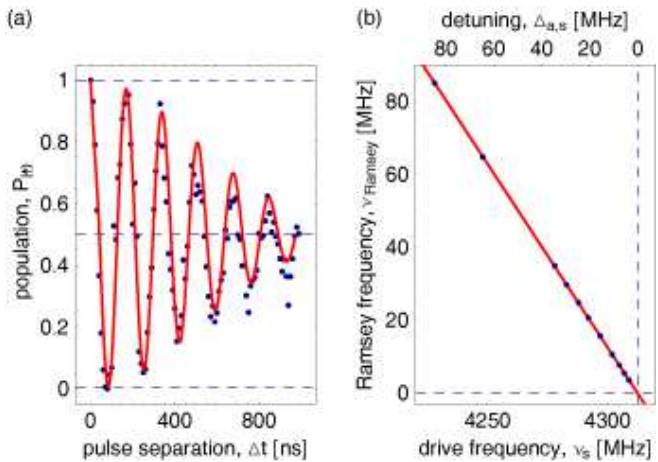}
\caption{(color online) (a) Measured Ramsey fringes (blue dots)
observed in the qubit population $P_{\left|\uparrow\right\rangle}$
\textsl{vs.} pulse separation $\Delta t$ using the pulse sequence
shown in Fig.~\ref{fig:setup}b and fit of data to sinusoid with
gaussian envelope (red line). (b) Measured dependence of Ramsey
frequency $\nu_{\rm{Ramsey}}$ on detuning $\Delta_{\rm{a,s}}$ of
drive frequency (blue dots) and linear fit (red line).}
\label{fig:Ramsey}
\end{figure}

We have determined the coherence time of the Cooper pair box from
a Ramsey fringe experiment, see Fig.~\ref{fig:setup}c, when biased
at the charge degeneracy point where the energy is first-order
insensitive to charge noise \cite{Vion02}. To avoid dephasing
induced by a weak continuous measurement beam \cite{schuster05} we
switch on the measurement beam only after the end of the second
$\pi/2$ pulse. The resulting Ramsey fringes oscillating at the
detuning frequency $\Delta_{\rm{a,s}} =
\omega_{\rm{a}}-\omega_{\rm{s}} \sim 6 \, \rm{MHz}$ decay with a
long coherence time of $T_2 \sim 500 \, \rm{ns}$, see
Fig.~\ref{fig:Ramsey}a. The corresponding  qubit phase quality
factor of $Q_{\varphi} = T_2 \omega_{\rm{a}}/2 \sim 6500$ is
similar to the best values measured so far in qubit realizations
biased at such an optimal point \cite{Vion02}. The Ramsey
frequency is shown to depend linearly on the detuning
$\Delta_{\rm{a,s}}$, as expected, see Fig.~\ref{fig:Ramsey}b. We
note that a measurement of the Ramsey frequency is an accurate
time resolved method to determine the qubit transition frequency
$\omega_a = \omega_s + 2\pi \, \nu_{\rm{Ramsey}}$.

In conclusion, performing Rabi and Ramsey experiments we have
observed high visibility in the oscillations of state population
of a superconducting qubit. The temporal response and the
back-action of the read-out are quantitatively understood and well
characterized.  Our charge qubit, which is embedded in a well
controlled electromagnetic environment, has $T_1$ and $T_2$ times
among the longest realized so far in superconducting systems. The
simplicity and level of control possible in this circuit QED
architecture makes it an attractive candidate for superconducting
quantum computation.

\begin{acknowledgments}
We would like to thank Michel Devoret and Jay Gambetta for
discussions. This work was supported in part by NSA and ARDA under
ARO contract number DAAD19-02-1-0045, and the NSF under grants
ITR-0325580 and DMR-0342157, the David and Lucile Packard
Foundation, the W.~M.~Keck Foundation, and the NSERC of Canada.
\end{acknowledgments}

\end{document}